\documentclass[russian]{rcdj}

\usepackage[times]{pscyr}
\usepackage{format}

\begin{document}






\setcounter{page}{265}
\journal{REGULAR AND CHAOTIC DYNAMICS, V.\,9, \No3, 2004}
\runningtitle{TWO-BODY PROBLEM ON A SPHERE}
\title{TWO-BODY PROBLEM ON
A SPHERE. REDUCTION, STOCHASTICITY, PERIODIC ORBITS}
\runningauthor{A.\,V.\,BORISOV, I.\,S.\,MAMAEV, A.\,A.\,KILIN}
\authors{A.\,V.\,BORISOV, I.\,S.\,MAMAEV, A.\,A.\,KILIN}
{Institute of Computer Science\\
Udmurt State University\\
Universitetskaya, 1, 426034, Izhevsk, Russia\\
E-mail: borisov@rcd.ru, mamaev@rcd.ru, aka@rcd.ru}

\abstract{We consider the problem of two interacting particles on a sphere. The
potential of the interaction depends on the distance between the
particles. The case of Newtonian-type potentials is studied in most
detail. We reduce this system to a system with two degrees of freedom and
give a number of remarkable periodic orbits. We also discuss integrability
and stochastization of the motion.}
\amsmsc{37N05, 70F10}
\doi{10.1070/RD2004v009n03ABEH000280}
\received 25.06.2004.

\maketitle

\renewcommand{\theparagraph}{\arabic{section}.\arabic{paragraph}}
\makeatletter
\@addtoreset{equation}{section}
\renewcommand{\theequation}{\thesection.\arabic{equation}}
\makeatother

\vspace*{-7mm}

\section{Introduction and historical notes}

The origins of the analysis of motion of mass particles and rigid bodies
in constant curvature spaces (plane, two- and three-dimensional spheres,
and Lobachevsky space) trace back to the classical XIXth century papers by
Serret, Killing, Lipschitz, Liebmann, Schering, etc. One of the most
comprehensive is the paper by W.\,Killing~\cite{bmk2-1}. The author
studied the motion of a particle in the ``central field'', a curved-space
analog of the Newtonian field (the Kepler problem). He also presented the
generalization of the Euler two-center problem and $N$-dimensional
analogs of the above-mentioned problems. He derived and partially studied
the equations of motion of a rigid body in these spaces and discussed some
issues of the theory of the Newtonian potential. More than a century after
that, the study of dynamics in constant curvature spaces attracted the
interest of scientists again. This resulted in appearance of a series of
papers, reviewed in~\cite{bmk2-9}. 
Note, though, that some of the classical results were independently
rediscovered in these papers.

One of the major distinctive features of curved spaces is the absence of
translational (Galilean) invariance, resulting in the absence of the
center-of-inertia integrals and non-existence of the corresponding
barycentric frame of reference. As a result, this takes us out of the
realm of the classical celestial mechanics.

For example, the two-body problem, when considered on a two-dimensional
(or three-dimensional) sphere~$S^2$~($S^3$) or on a Lobachevsky plane (or
space)~$L^2$ ($L^3$), can no longer be reduced to the corresponding
problem of motion in a central potential field (an analog of the Kepler
problem) and, generally, is not integrable, as will be shown below.
Hereinafter, we will discuss only the case of a two-dimensional
sphere~$S^2$, though the obtained results can be easily extended to~$L^2$
and their many-dimensional generalizations.

First of all, let us note that the Kepler problem of motion in a central
potential field on a sphere is integrable, and the orbits are ellipses
with one of the foci at the center. An analog of the Newtonian potential
on a sphere was first offered, apparently, by Serret in~\cite{bmk2-7}
(1860), the same for a Lobachevsky plane was done by Lobachevsky himself
and Bolyai. A generalization of the Bertrand theorem to a curved space was
performed by Liebmann (1903). Analogs of Kepler's laws for~$S^n$ and~$L^n$
can be found in a number of papers, of which the most fundamental is the
above-cited paper by W.\,Killing. Among the modern explorations of the
Kepler problem from various perspectives, we
recommend~\cite{bmk2-2,bmk2-3,bmk2-9,bmk2-14}.

So, an analog of the Newtonian potential on~$S^2$~($L^2$) is
\begin{equation}
\label{bmk2-eq-1}
U=-\gamma\ctg\theta\qquad (U=-\gamma\cth\theta),
\end{equation}
where $\theta$ is the longitude measured from the pole, at which the
gravitating center is located, and~$\gamma$ is the gravitational constant.
The potential~\eqref{bmk2-eq-1} can be obtained as a centrosymmetric
solution to the Laplace\f Beltrami equations for~$S^3$~($L^3$) or obtained
from an analog of Bertrand theorem
for~$S^2$~($L^2$)~\cite{bmk2-2}.\looseness=-1

In~\cite{bmk2-9}, we state the restricted problem of two bodies on a
sphere (or on~$L^2$), where one of the bodies circumscribes a great circle on
the sphere with constant velocity (i.\,e. moves freely along a geodesic),
while the other moves in its field (with the
potential~\eqref{bmk2-eq-1}) and does not affect the first body.
In~\cite{bmk2-3} (as well as in~\cite{bmk2-9}), we offer also a numerical
analysis of the Poincar\'e map for this problem, which is not integrable,
judging by its stochastic orbits; we also use the averaging method
to study the perigee motion caused by the curvature of the space.
In~\cite{bmk2-4}, it is proved, for the restricted problem, the non-existence
of an additional meromorphic analytic integral.

\section{Reduction of the two-body problem on~$S^2$}

The general (not restricted) two-body problem on~$S^2$~($L^2$) was studied
in respect to reduction in~\cite{bmk2-5}, where the quantization problems
were also discussed. The general method of reduction, offered
in~\cite{bmk2-5}, was first suggested by E.\,Cartan and further developed
by J.\,Marsden and A.\,Weinstein. This method is based on rather formal
argumentation, which, at the same time, has a profound differential
geometric interpretation. Unfortunately, the method does not always
gives and suitable
expressions, which have been sought after in the classical celestial
mechanics (for example, when reducing the three-body problem).

In this paper, to reduce the unrestricted two-body problem on a sphere, we
use Bour's reduction~\cite{bmk2-15} of the spatial three-body problem
 (similarly, Radau's reduction can be used~\cite{bmk2-15}). In
a sense, the reduction we need can be regarded as its special case. Connection
between the mentioned problems is based on a famous observation by Jacobi:
once a barycentric frame of reference was introduced, which is equivalent
to reduction to the center of inertia, the classical three-body problem
could be reduced to the problem of two mass points moving in the fixed
space~$Oxyz$ with a potential, which was a function of the points'
distances from the fixed center~$O$ and of the angle between their
radius-vectors.
Such a reduced system has a (vector) integral of angular momentum, and
Bour's reduction is an effective reduction of the system's order using
this integral (elimination of a node). With that, two more degrees of
freedom are eliminated. In the case of a sphere, the distances~$r_1,r_2$
between two points and the center are fixed, so in the Bour-reduced system
we only have to put~$r_1=r_2=R$, $\dot r_1=\dot r_2=0$ ($R$ is the
sphere's radius). The specified reduction is valid for the potential, which arbitrarily depends on the distance between the points
(the two-body problem is sometimes referenced as E.\,I.\,Kugushev's
problem~\cite{lit12}).

\fig<bb=0 0 81mm 70.2mm>{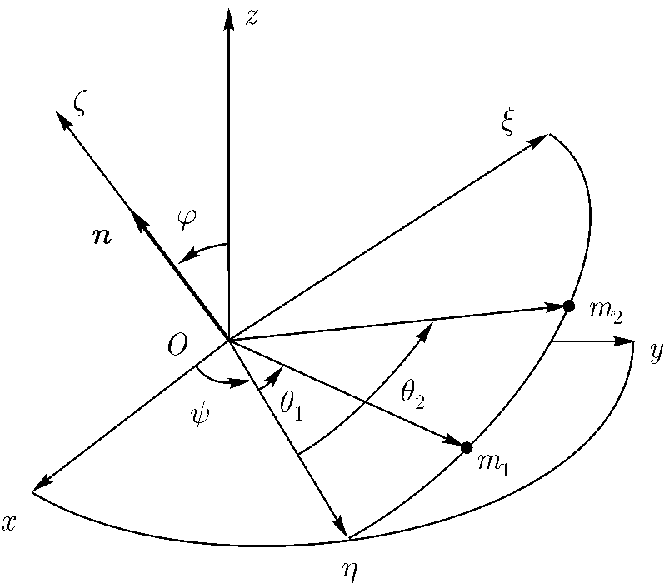}

Below we give the calculations in more detail. Let~$Oxyz$ be a fixed frame
of reference, and $O\xi\eta\zeta$~--- a moving frame of reference, such
that the plane~$O\eta\xi$ contains the mass points~$m_1$ and~$m_2$. The
axis~$\eta$, being an intersection of the plane~$(\eta\xi)$ with~$(xy)$,
is the line of nodes (see Fig.~\ref{mam1.eps}). The mass points are
described by the coordinates~$\theta_1$ and~$\theta_2$ (the angular polar
coordinates in the plane~$O\eta\xi$) and~$\vfi$,~$\psi$ (these angles
specify the location of the frame~$O\xi\eta\zeta$ relative to~$Oxyz$). It
is easy to see that all these variables are identical with the Euler
angles from the rigid body dynamics (this analogy was first noted by
J.\,Silvestr).

In terms of the introduced variables, the Cartesian coordinates read:
\begin{equation}
\label{bmk2-eq-2}
\begin{aligned}
x_i & = -R(\cos\vfi\cos\psi\sin\theta_i+\sin\psi\cos\theta_i),\\
y_i & = -R(\cos\vfi\sin\psi\sin\theta_i-\cos\psi\cos\theta_i),\\
z_i & = R\sin\theta_i\sin\vfi,
\end{aligned}\quad
i=1,2.
\end{equation}

Using \eqref{bmk2-eq-2}, the Lagrangian function of the system can be
written as
\begin{multline}
\label{bmk2-eq-3}
\cL=\frac{R^2}{2}\Bigl(m_1(\dot \theta_1+\dot\psi\cos \vfi)^2+
m_2(\dot \theta_2+\dot\psi\cos\vfi)^2
+J\dot\vfi^2+\\
+R\sin^2\vfi\dot\psi^2-2L\dot\vfi\dot\psi\sin\vfi)-U(\theta_1-\theta_2)\Bigr),
\end{multline}
where
$$
\begin{gathered}
J = m_1\sin^2\theta_1+m_2\sin^2\theta_2,\q
K = m_1\cos^2\theta_1+m_2\cos^2\theta_2,\\
L = m_1\sin\theta_1\cos\theta_1+m_2\sin\theta_2\cos\theta_2.
\end{gathered}
$$
\goodbreak

Let the axis $Oz$ be directed along the vector of angular
momentum~$|\bM|=c$. The Lagrangian equations with the
Lagrangian~\eqref{bmk2-eq-3} admit the integral
\begin{equation}
\label{bmk2-eq-4}
\pt{\cL}{\dot\psi}=M_z=c=\const
\end{equation}
and invariant relations
\begin{equation}
\label{bmk2-eq-5}
\pt{\cL}{\dot\vfi}=M_\eta=0,\quad \pt{\cL}{\vfi}=M_{z\times\eta}\dot \psi=0.
\end{equation}

Indeed,
$$
\dot M_\eta=M_{z\times\eta}\dot\psi,\quad \dot
M_{z\times\eta}=-M_\eta\dot\psi,
$$
where~$M_z$, $M_\eta$, $M_{z\times\eta}$ are the projections of the
angular momentum~$\bM$ onto the given axes~\cite{bmk2-15} (see
Fig.~1). The invariant manifold~\eqref{bmk2-eq-4},~\eqref{bmk2-eq-5} is
explicitly given by:
\begin{equation}
\label{bmk2-eq-5a}
\begin{gathered}
\dot\psi=\frac{cJ}{m_1m_2R^2\sin^2(\theta_1-\theta_2)},\q
\dot\vfi=\frac{cL\sin\vfi}{m_1m_2R^2\sin^2(\theta_1-\theta_2)},\\
\cos\vfi=\frac{m_1\dot\theta_1+m_2\dot\theta_2}{c/R^2-(m_1+m_2)\dot\psi}.
\end{gathered}
\end{equation}

It follows from \eqref{bmk2-eq-4} that the variable~$\psi$ is cyclic, so
we can reduce the system's order by one degree of freedom. Besides, the
variable~$\vfi$ is also cyclic on the invariant
manifold~\eqref{bmk2-eq-5}, and here we again can reduce the number of
degrees of freedom by one, using the Routh reduction procedure. The Routh
function is
\begin{equation}
\label{bmk2-eq-6}
\begin{gathered}
\cR=\cL{\,-\,}\pt{\cL}{\dot\psi}\dot\psi{\,-\,}\pt{\cL}{\dot\vfi}\dot\vfi{\,=\,}
\frac{R^2}{2}(m_1\dot\theta_1^2{\,+\,}m_2\dot\theta_2^2){\,-}\\
{-\,}\frac{R^2J(m_1\dot\theta_1{+}m_2\dot\theta_2)^2}{2(J(m_1{+}m_2){-}
m_1m_2\sin^2(\theta_1{-}\theta_2))}{\,-\,}\frac{c^2J}{2m_1m_2R^2\sin^2
(\theta_1{-}\theta_2)}{-}U(\theta_1{-}\theta_2).
\end{gathered}
\end{equation}
Using the Legendre transformations
\begin{equation}
\label{bmk2-eq-6a}
\begin{gathered}
p_{i}=\pt{\cR}{\dot\theta_i}=m_iR^2\dot\theta_i-\frac{m_iR^2J(m_1\dot\theta_1+m_2\dot\theta_2)}
{J(m_1+m_2)-m_1m_2\sin^2(\theta_1-\theta_2)},\q i=1,\,2,\\
\end{gathered}
\end{equation}
we obtain the equations of motion of the Hamiltonian form with the
Hamiltonian
\begin{equation}
\label{bmk2-eq-7}
\cH=\frac{1}{2R^2}\Bigl(\frac{p_{1}^2}{m_1}+\frac{p^2_{2}}{m_2}+
(c^2-(p_{1}+p_{2})^2)\frac{J}{m_1m_2\sin^2(\theta_1-\theta_2)}\Bigr)+
U(\theta_1-\theta_2).
\end{equation}
This is the sought-for reduced system with two degrees of freedom.

The absolute motion (i.\,e. motion in the fixed frame of reference) can be
obtained, using the two additional quadratures given
by~\eqref{bmk2-eq-5a}.

\begin{rem*}
On the manifold \eqref{bmk2-eq-5a}, the domain of possible values of the new
variables~$p_1$ and~$p_2$ is bounded because it follows
from~\eqref{bmk2-eq-5a} that~$\cos\vfi=\frac{p_1+p_2}{c}$ and $|\cos\vfi|\le 1$.
\end{rem*}

So, the phase space of the reduced system is parametrized with the
canonical variables~$\ta_1^{},\,\ta_2^{},\,p_1^{},\,p_2^{}$. The angle
variables~$\ta_1^{},\,\ta_2^{}\bmod 2\pi$ define a two-dimensional torus
(configuration space of the reduced system), while the conjugate
momenta~$p_1^{}$,~$p_2^{}$ form the strip~$|p_1^{} + p_2^{}| = |c\cos
\varphi| \le |c|$. Besides, it is necessary to identify the points of the
phase space with coordinates~\eq[kor]{
(\ta_1^{},\,\ta_2^{},\,p_1^{},\,p_2^{}) \q \t{and}\q (\pi -
\ta_1^{},\,\pi-\ta_2^{},\,-p_1^{},\,-p_2^{}). }

Indeed, according to~\eqref{bmk2-eq-2}, the different sets of coordinates,
$(\ta_1^{},\,\ta_2^{},\,\varphi,\,\psi)$ and~$({\pi - \ta_1^{}},\pi-
\ta_2^{},\,\pi-\varphi_1^{},\,\pi + \psi)$, define the same configuration
of the bodies on the sphere. For the velocities we have~$(\pi -
\ta_1^{})\,\dot{} = -\dot \ta_1^{}$, $(\pi - \ta_2^{})\,\dot{} = -\dot
\ta_2^{}$ and, using~\eqref{bmk2-eq-6a}, obtain~\eqref{kor}. We believe it
would be an interesting to develop a better geometric
description of the reduced phase space.

The Hamiltonian~\eqref{bmk2-eq-7} is a homogeneous, but not positively
defined, quadratic function of the momenta. Since the
system~\eqref{bmk2-eq-7} is not natural, well-developed methods of
topological and qualitative analysis~\cite{lit18} can hardly be applied to
it.

\section{Particular solutions and the Smale diagram of the two-body
problem on~$S^2$}

Let us start with the general problem of $n$ interacting bodies. The
potential of interaction~$U$ depends on the distances between the bodies.

A particular solution to the~$n$-body problem will be called a {\it
rigid-body motion} if the distances between the bodies remain constant. A
particular solution to the~$n$-body problem will be called a {\it
stationary configuration {\rm (}relative equilibrium}\/) if all the bodies
rotate uniformly, with the same angular velocity, about the axis
containing the vector~$\bM$.

It is obvious that any stationary configuration is a rigid-body motion,
but the reverse is, generally, not true.

\subsection{Stationary configurations and the Smale diagram}

The stationary configurations in the general~$n$-body problem on~$S^2$ can
be obtained as the critical points of the reduced potential.

\setcounter{pro}{0}
\begin{pro}\label{19pro1}
The stationary configurations {\rm(}relative equilibriuma\/{\rm)} of
the~$n$-body problem correspond to the critical points of the reduced
{\rm(}effective\/{\rm)} potential~\eq[eq1]{ U_\ast = U - \frac{1}{2}
\omega^2 I = U - \frac{\bM^2}{2{I}}, } where $\om$ is the angular velocity
of the configuration,~$I$ is the total moment of inertia of the system
with respect to the axis of rotation, $\bM^2$ is the squared total angular
momentum of the system of bodies in a fixed frame of reference.
\end{pro}

\proof Consider in the frame of reference, rotating with the angular
velocity~$\omega$ about the axis~$z$. In this frame, the Lagrangian (in
terms of spherical coordinates) reads: \eq[eq2]{ {\cL} = \frac{1}{2}\suml
m_i^{} R^2(\dot \ta_i^2 + \sin^2 \ta _i^{} \dot \varphi_i^2) + \omega
\suml m_i^{} \dot \varphi_i^{} R^2 \sin^2 \ta_i^{} - U + \frac{1}{2}
\omega^2 I, } where $I = \suml m_i^{} R^2 \sin^2 \ta_i^{}$ is the total
moment of inertia with respect to the~$z$-axis ($i$ is the number of a
particle).

Then, using equations of motion with the Lagrangian~\eqref{eq2} and
bearing in mind that~$\dot \ta_i^{} = \dot \varphi_i^{} = 0$ (stationary
solution), we obtain the following conditions of the relative equilibrium:
\eq[eq3]{ \pt{}{\ta_i^{}}\left(U - \frac{1}{2}\omega^2 I\right) = 0,\qq
\pt{}{\varphi_i^{}}\left(U - \frac{1}{2}\omega^2 I\right) = 0,\qq i =
1\dts n. }

To finish the proof, we have yet to express the angular momentum of the
stationary configuration in the fixed frame of reference: \eq*{ M_x^{} =
M_y^{} = 0,\qq M_z^{} = \suml m_i^{} R_i^2 \sin^2 \ta_i^{} \dot
\varphi_i^{} = \omega \suml m_i^{} R^2 \sin^2 \ta_i^{} = I\omega. \qed}

Now we consider the case of the generalized Newtonian potential and give a
complete description of the relative equilibriuma of the two-body problem
on~$S^2$. According to Proposition~\ref{19pro1}, it is necessary to find
the critical points of the reduced potential~\eq*{ U_\ast = -\text{\ae} R^2
m_1^{} m_2^{} \ctg \ta_{12} - \frac{1}{2} \omega^2 R^2 \left(m_1^{} \sin^2
\ta_1^{} + m_2 \sin^2 \ta_2\right)\!, } where $\ta_{12}$ is the angular
distance the particles on the sphere, and~$\text{\ae}$ is the gravitational
constant.

It can be shown that for relative equilibriuma of two particles on~$S^2$,
the following two observations are true:

\begin{enumerate}
\item {\it While moving, the particles never leave the great circle that
contains the axis of rotation\/}.

\item {\it When~$\text{\ae}>0$ {\rm(}the case of attraction\/{\rm)} the particles
stay on the great circle, with the axis of rotation between them\/ (see Fig.~\ref{_1.eps})}.
\end{enumerate}

Indeed, the cosine of the angle between the mass points can be rewritten
as \eq*{ \xi = \cos \ta_{12} = \cos \ta_1^{} \cos \ta_2 + \sin \ta_1^{}
\sin \ta_2^{} \cos (\varphi_1 - \varphi_2), } and we find that \eq*{
\pt{U_\ast}{\varphi_1^{}} = -\pt{U_\ast}{\varphi_2^{}} = \frac{\text{\ae} R^2
m_1^{} m_2^{}} {(1 - \xi^2)^{3/2}}\sin \ta_1^{} \sin \ta_2^{}
\sin(\varphi_1^{} - \varphi_2^{})= 0. } Hence, either~$\varphi_1^{} =
\varphi_2$, or~$\varphi_2^{} = \varphi_1^{} + \pi$. If~$\varphi_1 =
\varphi_2$ the equations~$\pt{U_\ast}{\ta_i^{}} = 0$ do not have a common
solution.

So, to find stationary configurations, it is necessary to find the
critical points of the function \eq*{ U_\ast = - \text{\ae} R^2 m_1^{} m_2^{}
\ctg(\ta_1^{} + \ta_2^{}) - \frac{1}{2} \omega^2 R^2 (m_1^{}
\sin^2\ta_1^{} + m_2^{} \sin^2 \ta_2^{}) } where the angles $\ta_1^{}$ and
$\ta_2^{}$ are measured from the axis of rotation (Fig.~\ref{_1.eps}).
These angles are given by \eq[eq4]{ \omega^2 \sin \ta_1^{} \cos \ta_1 -
\frac{\text{\ae} m_2^{}}{\sin^2(\ta_1^{} + \ta_2^{})}= 0, \qq \omega^2 \sin
\ta_2^{} \cos \ta_2 - \frac{\text{\ae} m_1^{}}{\sin^2(\ta_1^{} + \ta_2^{})}= 0. }

\fig<bb=0 0 63.7mm 63.7mm>{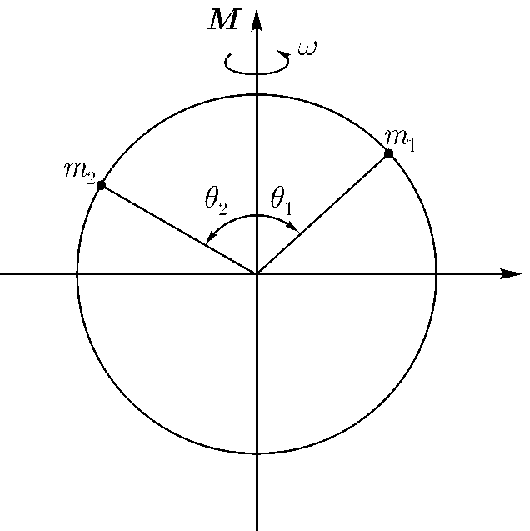}

From this, we obtain the equation that generalizes the planar ``law of the
lever'' and relates the angles $\ta_1^{}$ and~$\ta_2^{}$ \eq[eq5]{ m_1^{}
\sin 2 \ta_1^{} = m_2^{} \sin 2 \ta_2^{}. } Using~\eqref{eq4} and
\eqref{eq5}, one can show that for~$\text{\ae} > 0$, either~$\ta_1,\,\ta_2^{} <
{\pi}/{2}$, or~$\ta_1,\,\ta_2^{} > {\pi}/{2}$. These configurations are
identical up to the reflection about the equatorial plane. Therefore, we
will assume that \eq[eq6]{ 0 < \ta_1^{},\,\ta_2^{} < \frac{\pi}{2}. }

If we put $m_1^{} < m_2^{}$, then the equation has two roots on the
interval~\eqref{eq6} \vskip-0.5mm\noindent \eq[eq7]{ \ta_2^{(1)} =
\frac{1}{2}\arcsin\left(\frac{m_1^{}}{m_2^{}}\sin 2 \ta_1^{}\right)\!,\q
\ta_2^{(2)} = \frac{\pi}{2} -
\frac{1}{2}\arcsin\left(\frac{m_1^{}}{m_2^{}}\sin 2 \ta_1^{}\right)\!; }
\vskip-0.5mm\noindent here $\ta_2^{(1)} < \ta_1^{}$, and $\ta_2^{(2)} >
\ta_1$.

The first solution~$\ta_2^{(1)}$ describes a configuration, where the
heavier particle moves along the smaller circle. As~$R \to \fy$, this
solution tends to ordinary stationary configuration of the two-body
problem on a plane.

The second solution~$\ta_2^{(2)}$ describes a configuration, where the
heavier particle moves along the larger circle; this solution does not
have an analog in the two-body problem on a plane.\looseness=-1

\begin{figure}[!hb]
\centering\cfig<bb=0 0 129.8mm 45.7mm>{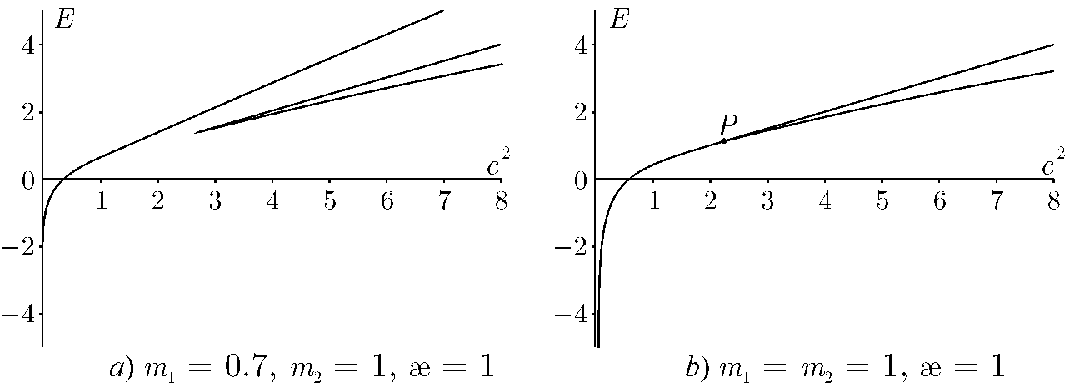} \caption{Bifurcation diagram
for the two-body problem with the gravitational potential~$U = -\text{\ae} m_1^{}
m_2^{} \ctg \ta$} \label{3.eps}\vspace{-3mm}
\end{figure}

Let us plot a bifurcation diagram (see~\cite{lit17}) for the two-body
problem on~$S^2$ on the integral plane~$(c^2 = \bM^2/R^4,\,E = h/R^2)$,
where~$h$ is the energy integral (Fig.~\ref{3.eps}). When~$m_1^{} \ne
m_2^{}$, we obtain two curves, corresponding to the solutions
of~\eqref{eq7} (see Fig. \ref{3.eps}{\it a}\/). When~$m_1^{} = m_2^{}$,
the curves merge at a point~$P$ (Fig. \ref{3.eps}{\it b}); in this case,
also according to~\eqref{eq7}, we have either~$\ta_1^{} = \ta_2^{}$,
or~$\ta_1^{} + \ta_2^{} = \pi/2$.\looseness=-1

Here we also give a bifurcation diagram for the two-body problem on a
sphere in the presence of a potential that, unlike the gravitational
potential, has no singularities (see Fig.~\ref{2.eps}). For that
 potential, in the paper~\cite{lit12} a geometrical
(topological) analysis of the integral manifolds~$\bM^2 = \const$, $h =
\const$ and of the level surface of the reduced system's energy integral is performed.
Note that, due to singularities at the poles in the case of the
gravitational potential~\eqref{bmk2-eq-1}, it is impossible to directly
apply the results of~\cite{lit12} to this case. So, the complete
topological analysis of three-dimensional isoenergetic manifolds does not
seem to have been done.

\fig<bb=0 0 129.5mm 52.3mm>{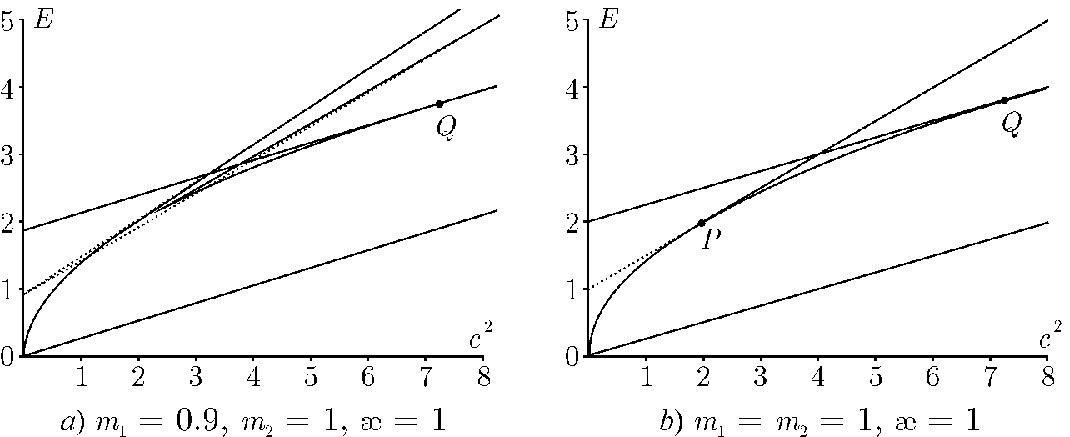}[Bifurcation diagram for the two-body
problem on a sphere with the potential~$U = \text{\ae} m_1^{} m_2^{}(1 - \cos
\ta)$]

\subsection{A particular solution in the case of equal masses}


The $n$-body problem on a plane has a particular solution (collinear
configuration), in which all the particles move along ellipses, while
staying at all times on the same straight line. This solution was first
offered by Euler (for the three-body problem) and Moulton (for
the~$n$-body case)~\cite{lit13}. In~\cite{lit13}, Moulton also proved that
for the case when all masses are different the number of
such collinear configurations (for different permutations of
the particles) is equal to~$\frac{n!}2$. Let us show that analogous
(pulsating) solutions on a sphere do not, generally, exist.

\begin{pro}
For the two-body $(n$-body\/{\rm)} problem on~$S^2$, when~$m_i^{} \ne
m_j^{}$, there are no solutions other than a stationary configuration, in
which all the particles stay on the same great circle that contains the
axis of rotation.
\end{pro}

\proof Let $Oz$ be the axis of rotation. The equations of motion in
spherical coordinates are
\eq[eq8]{
\begin{gathered}
m_i^{} \ddot \ta_i^{} = m_i^{} \sin \ta_i^{} \cos \ta_i^{} \dot
\varphi_i^2 - \pt{U}{\ta_i^{}}\\[-5pt]
\frac{d}{dt}\left( m_i^{} \sin^2 \ta_i^{} \dot \varphi_i^{}\right) = -
\pt{U}{\varphi_i^{}} = \suml_{j\ne 1,\,j \ne 1}^n U_{\xi_{ij}}' \sin
\ta_i^{} \sin \ta_j^{} \sin(\varphi_i^{} - \varphi_j^{}),
\end{gathered}
} where $\xi_{ij}^{} = \cos \ta_i^{} \cos \ta_j^{} + \cos \ta_i^{} \sin
\ta_j^{} \cos (\varphi_i^{} - \varphi_j)$.
\goodbreak

If the particles always stay on a great circles that contains the axis of
rotation, then either~$\varphi_i^{} = \varphi_j^{}$, or~$\varphi_j^{} =
\varphi_i^{} + \pi$ and, besides, \eq*{ \dot \varphi_i^{} = \dot
\varphi_j^{} = \dot \psi, \qq i,\,j = 1\dts n. } Therefore, according
to~\eqref{eq8}, the following must hold: \eq[eq8.25]{ m_i^{}
\sin^2\ta_i^{} \dot \psi = m_i^{} c_i^2,\qq c_i^{} = \const }

Hence, all the angles~$\ta_i$ depend only on the constants~$c_i^{}$ and
the function~$\lm^2(t) = 1/\dot \psi(t)$ \eq[eq8.5]{ \zeta_i^{} = \sin
\ta_i^{} = c_i^{} \lm(t). } Differentiating this and substituting into the
first equation of~\eqref{eq8}, we obtain \eq[eq9]{ \frac{m_i^{}
c_i^{}}{\sqrt{1-c_i^{} \lm^2}}\left(\ddot \lm + \frac{c_i^2 \lm \dot
\lm^2}{1 - c_i^2 \lm^2}\right) = m_i^{} c_i^{} \lm^{-3}\sqrt{1-c_i^2
\lm^2} - \pt{U}{\ta_i^{}} } where $\pt{U}{\ta_i^{}} =
\pt{U}{\zeta_i^{}}\pt{\zeta_i^{}}{\ta_i^{}} = \sqrt{1 - c_i^2 \lm^2}
\pt{U}{\zeta_i^{}}$. The required solution exists if~\eqref{eq9} can be
made independent on the particle's number~$i$. Provided that~$\dot \lm
\not\equiv 0$, this can be done if~$\vartheta_i^{}$ are assumed to be the
critical points of the reduced potential~\eqref{eq1} and all~$c_i^{}$ are
assumed to be equal, i.\,e. \eq{ \pt{U}{\ta_i^{}} = \omega^2 m_i^{} \sin
\ta_i^{} \cos \ta _i^{}, \qq c_i^{} = c = \const. } Then, taking into
account~\eqref{eq8.5} and~\eqref{eq9}, incompatible systems are obtained
\eq*{ m_i^{} \sin 2 \ta_i^{} = \pm m_j^{} \sin 2 \ta_j^{} \q \t{and} \q
\sin \ta_i^{} = \pm \sin \ta_j^{}, } which does not have a solution~$m_i^{}
\ne m_j^{}$. \qed

Let us consider more closely the case of equal masses in the two-body
problem,~$m_1^{} = m_2^{} = 1$ (with~$\varphi_2^{} = \varphi_1^{} + \pi$
and~$\ta_1^{} = \ta_2^{} = \ta$). Using directly~\eqref{eq8}
and~\eqref{eq8.25}, we obtain \eq[eq21_5]{
\begin{gathered}
\dot \psi = \dot \varphi_1^{} = \dot \varphi_2^{} = \frac{c}{\sin^2 \ta}\qq
\ddot \ta = \sin \ta \cos \ta \dot \psi^2 - \frac{\text{\ae}}{\sin^2 2\ta},
\qq c = \const.
\end{gathered}
} These equations describe the motion of a particle on a sphere in the
field of an attracting center with the potential \eq*{ U = - \frac{1}{2}
\text{\ae} \ctg 2\ta. }

For this potential, the orbits are unclosed, which showes another difference
between curved and flat spaces. It is known that, due to uniformity
condition, the particles' orbits in flat spaces are ellipses with one of
the foci at the center of mass.

Nevertheless, orbits on a sphere are closed in some rotating frame of
reference. Moreover, one can choose the velocity of rotation so that both
particles will move along one and the same curve (such motion is called a
\emph{relative choreography}~\cite{lit15,lit14}, see Fig.~\ref{chor.eps},
$B_1$, $B_2$, $B_3$).

\subsection{Rigid-body motions in the two-body problem}

Consider the motion of particles in a flat space~$\mR^3$ in a fixed frame
of reference. In this case, the relative equilibriuma of the two-body
($n$-body) problem in the center-of-mass reference frame, generally
(i.\,e. when the velocity of the center of mass is not zero), correspond
to the simplest rigid-body solutions. The mutual distances between the
particles are always the same. We show that in a curved space (on~$S^2$), lacking the notion of
the center of mass, all the rigid-body solutions of the two-body problem
are stationary configurations discussed above~\eqref{eq4},~\eqref{eq7}.
Therefore, the corresponding family of configurations has less free
parameters than such a family in a flat space.

\begin{teo*}
When the interaction potential depends only on the mutual distances, all
rigid-body solutions of the two-body problem on~$S^2$, are stationary
configurations.
\end{teo*}

\proof The proof is based on successive differentiation of the invariant
relation \eq[eq10]{ (\bx,\,\by) = \cos \ta_0^{} = \const, } where
$\bx,\,\by$ are the radius-vectors of the particles on~$S^2$
in~$\mathbb{R}^3$, and finding the maximal invariant manifold that
contains these motions.

Now let us perform an algebraic reduction of our system by restricting it
onto a level surface of~$\bM$. For the variables \eqa*{ \bM = m_1^{} \dot
\bx \x \bx + m_2^{} \dot \by \x \by,\qq \bL = \mu(\dot \bx \x \bx - \dot
\by \x \by), } where $\mu = \frac{m_1^{}m_2^{}}{m_1^{} + m_2^{}}$, we have
nine equations:\vspace{-2mm}
\eq[eq11]{
\begin{gathered}
\dot \bL = U_\xi' \bx \x y\\
\dot \bx = \frac{1}{m_1^{} + m_2^{}} \bx \x \bM + \frac{1}{m_1^{}} \bx \x
\bL,\\ \dot \by = \frac{1}{m_1^{} + m_2^{}} \by \x \bM - \frac{1}{m_1^{}}
\by \x \bL.
\end{gathered}
} Here $U(\xi)$ is the potential energy of the interaction, expressed in
terms of the cosine of the angle between the particles~$\xi =
(\bx,\,\by)$, and vector~$\bM$ is a parameter. Due to the condition~$(\dot
\bx,\,\bx) = (\dot \by,\,\by) = 0$ and definition of~$\bx,\,\by$, the
following relations hold: \eq[eq12]{
\begin{gathered}
\bx^2 = \by^2 = 1;\\
\frac{m_1}{m_1^{} + m_2^{}}(\bx,\,\bM) + (\bx,\,\bL) = 0, \q
\frac{m_2^{}}{m_1^{} + m_2^{}}(\by,\,\bM) - (\by,\,\bL) = 0.
\end{gathered}
} The energy integral in this case reads \eq*{ E = \frac{1}{2 \mu} \bL^2 +
U(\xi) = \const. }

Differentiating the relation~$(\bx,\,\by) = \const$
by time and using~\eqref{eq11} and~\eqref{eq12}, we get
\eq[eq5_26]{
\begin{gathered}
(\bx,\,\by \x \bL) = 0,\\
\frac{1}{\mu}\bL (\bx,\,\by) + U'(\xi) (\bx \x \by)^2 +
\frac{2}{m}(\bx,\,\bM)(\by,\,\bM) = 0,\\
(\bx,\,\bM \x \bL) = 0.
\end{gathered}
}

It follows from the second and the fourth equations of~\eqref{eq5_26} that
the vectors~$\bx,\,\by,\,\bM$, and~$\bL$ belong to the same plane.
Express~$\bM,\,\bL$ in terms of~$\bx,\,\by$: \eq[eq5_27]{
\begin{gathered}
\bM = \frac{(\bM,\,\bx) - \cos \ta_0^{} (\bM,\,\by)}{\sin^2 \ta_0^{}}
\bx + \frac{(\bM,\,\by) - \cos \ta_0^{}(\bM,\,\bx)}{\sin^2 \ta_0^{}}\by,\\
\bL = \frac{(\bL,\,\bx) - \cos \ta_0^{} (\bL,\,\by)}{\sin^2 \ta_0^{}} \bx
+ \frac{(\bL,\,\by) - \cos \ta_0^{}(\bL,\,\bx)}{\sin^2 \ta_0^{}}\by;
\end{gathered}
} substituting this into~\eqref{eq11} and \eqref{eq5_26}, we have
$(\bM,\,\bx) = \const$, and~$(\bM,\,\by) = \const$. Thus, the particles
move in circles about some axis defined by the vector~$\bM$, and during
the motion the particles and the axis always lie in the same plane.\qed

\subsection{The case of $\bM = 0$ and the two-body problem on a circle.
Collisions}

In the two-body problem on a sphere (with an arbitrary potential,
depending only on the inter-particle distance), the motion with zero total
momentum ($\bM = 0$) is a special case. We will show that the particles in
this case move along a great circle, which holds its position in a fixed
frame of reference.

Indeed, for $\bM = 0$, equations~\eqref{eq11} read \eq[eq14]{ \dot \bx =
\frac{1}{m_1^{}} \bx \x \bL,\qq \dot \by = - \frac{1}{m_2^{}} \by \x
\bL,\qq \dot \bL = U_\xi' \bx \x \by, } while the relations \eqref{eq12}
take the form \eq[eq15]{ (\bx,\,\bL) = (\by,\,\bL) = 0. }

Calculating the orbital derivative of the normal to the particles'
plane~$\bn = \frac{\bx \x \by}{|\bx \x \by|}$ under the flow
of~\eqref{eq14} and using~\eqref{eq15}, we get~$\dot \bn \equiv 0$. Thus,
the plane that contains the radius-vectors of the particles is fixed,
and the particles move along a great circle.

\begin{rem*}
In the two-body problem on a plane, when the momentum is zero, the
particles move along the same straight line, fixed in the center-of-mass
reference frame. This frame moves uniformly and in a straight line.
\end{rem*}

Since for zero momentum~$\bM = 0$, the two-body problem on a sphere is
reduced to a system on a circle, we will consider the latter problem in
more detail (for arbitrary~$\bM$). We will also show that in this case
there is a center-of-mass reference frame, in which the behavior of the
particles does not depend on the value of~$\bM$. The new variables are
\eq{ \psi = \frac{m_1^{} \ta_1^{} + m_2^{} \ta_2^{}}{m_1^{} + m_2^{}},\qq
\theta = \ta_1^{} - \ta_2^{}, } where $\ta_1^{}$ and~$\ta_2^{}$ are the
angles that define the positions of the particles on the circle. So, the
Lagrangian reads \eq*{ \cL = \frac{1}{2} (m_1^{} + m_2^{})\dot \psi^2 +
\frac{1}{2}\mu \dot \ta^2 - U(\ta),\qq \mu = \frac{m_1^{}m_2^{}}{m_1^{} +
m_2^{}}. } This implies that the ``center-of-mass angle''~$\psi$ varies
uniformly with time: \eq{ \dot \psi = \frac{|\bM|}{R^2 (m_1^{} + m_2^{})}
= \const. } The time dependency of~$\ta$ is given in terms of quadratures
\eq[eq16]{ \intl_{\ta_0^{}}^{\ta} \frac{d\ta}{\sqrt{2\mu^{-1}(h_1^{} -
U(\ta))}} = t - t_0^{}, } where $\ta_0^{},\,t_0^{}$, and~$h_1^{}$ are some
constants.\goodbreak

For the generalized Newtonian potential~$U(\ta) = - \text{\ae} m_1^{} m_2^{} \ctg
\ta$, we find that the particles will collide in finite time. If~$t_0^{} =
0$ at the time of collision, the time dependency of the angle can be
written in the form of the Puiseux series \eq[eq17]{ \ta(t) = t^{2/3}
\suml_{n=0}^\fy c_n^{} t^{n/3}. } Here $c_0^{} \ne 0$, while the
coefficients $c_n^{}$ with odd indices are zero ($c_{2n+1} = 0$). In other
words,~$\ta(t)$ is an even function of time. Consequently, the particles
undergo perfectly elastic collision (for a suitable
regularization).

\begin{rem}
The proof of \eqref{eq17} is based on the analysis of~\eqref{eq16}
near~$\ta = 0$. Introducing a new variable~$x = \tg \ta$, we
express~\eqref{eq16} in the following form (for~${t_0^{} = 0}$):\vspace{-2mm}
\eq*{
\intl_0^x \frac{\sqrt{x} dx}{(1+ x^2)\sqrt{1+ax}} = \zeta,\qq \zeta =
\sqrt{2\text{\ae}(m_1^{} + m_2^{})}t,\qq a =
\frac{h_1^{}}{\text{\ae} m_1^{} m_2^{}}.\vspace{-2mm}
}
Using the Taylor expansion of the integrand, we obtain the following
series equations defining $x(t)$:\vspace{-2mm}
\eq*{
u^3 \suml_{n=0}^{\fy} g_n^{}
u^{2n} = \zeta,\qq u^2 = x,\vspace{-2mm}
}
where $g_n^{}$ are constants that depend
on~$a$. After eliminating~$u$, we have\vspace{-2mm}
\eq*{
x(\zeta) = \zeta^{2/3}
\suml_{n=0}^\fy b_n^{} \zeta^{2n/3}.\vspace{-2mm}
}
\end{rem}

\begin{rem}
We believe it would be interesting to generalize the classical results
obtained by Zigel, concerning zero measure of collision orbits, to this
problem (i.\,e. to specify the collision manifold).

Let us also mention a long-discussed question~\cite{bmk2-9,bmk2-14} about
a possibility of extending Sundman's results (i.\,e. solutions in power
series) to the two- and three-body problems in curved spaces.
Unfortunately, the answer to this question has not yet been found and is,
most likely, negative.
\end{rem}

\section{The Poincar\'e section. Numerical analysis}

\begin{figure}[!h]
\centering\cfig<width=0.75\textwidth>{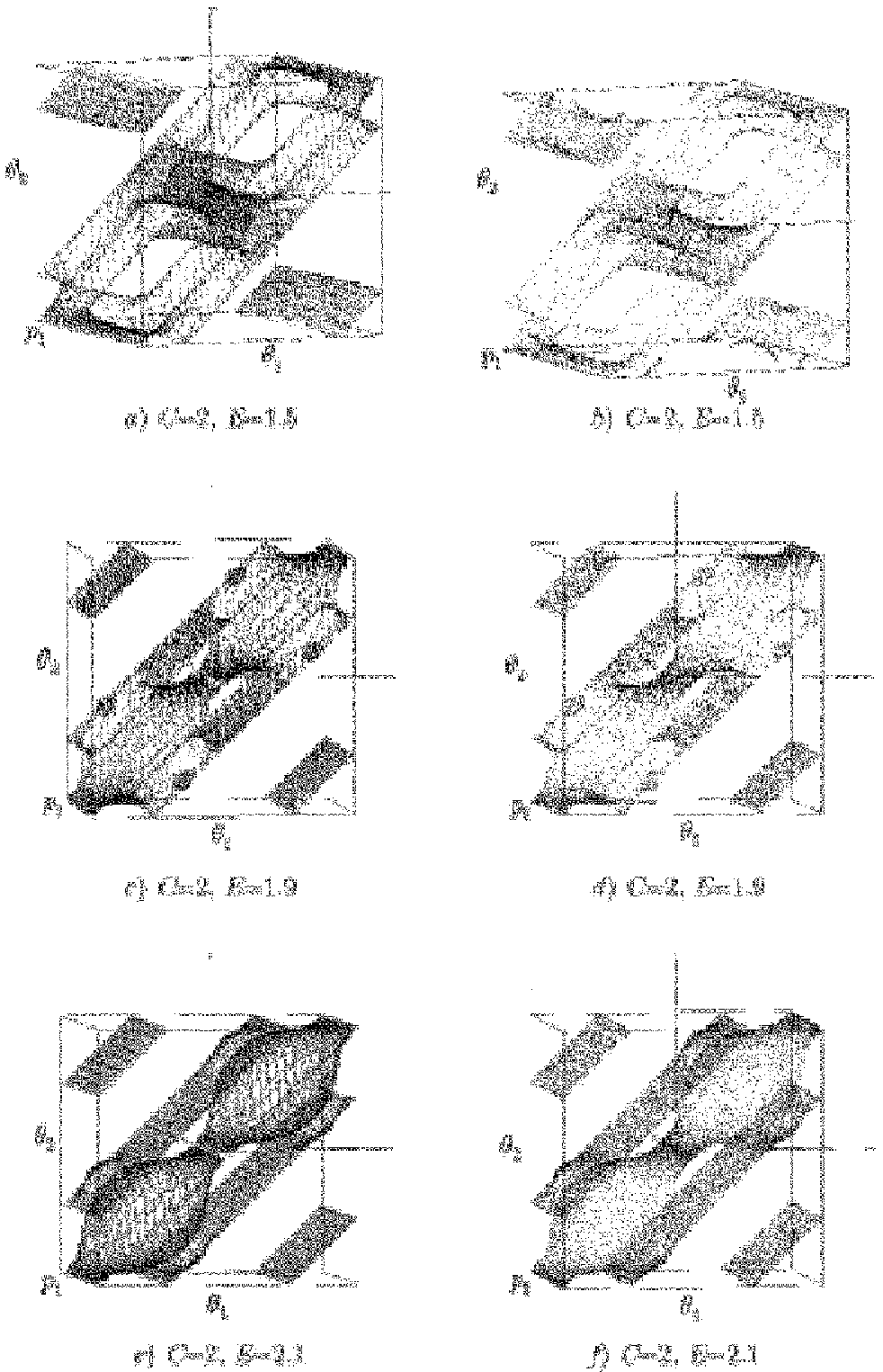}
\smallskip
\caption{Energy level sections ({\it a, c, e, g, k}\/) and corresponding
Poincar\'e maps ({\it b, d, f, h, l}\/) for different values of~$E$
and~$c$.} \label{pic1}
\end{figure}
\addtocounter{figure}{-1} \fig<width=0.6\textwidth>{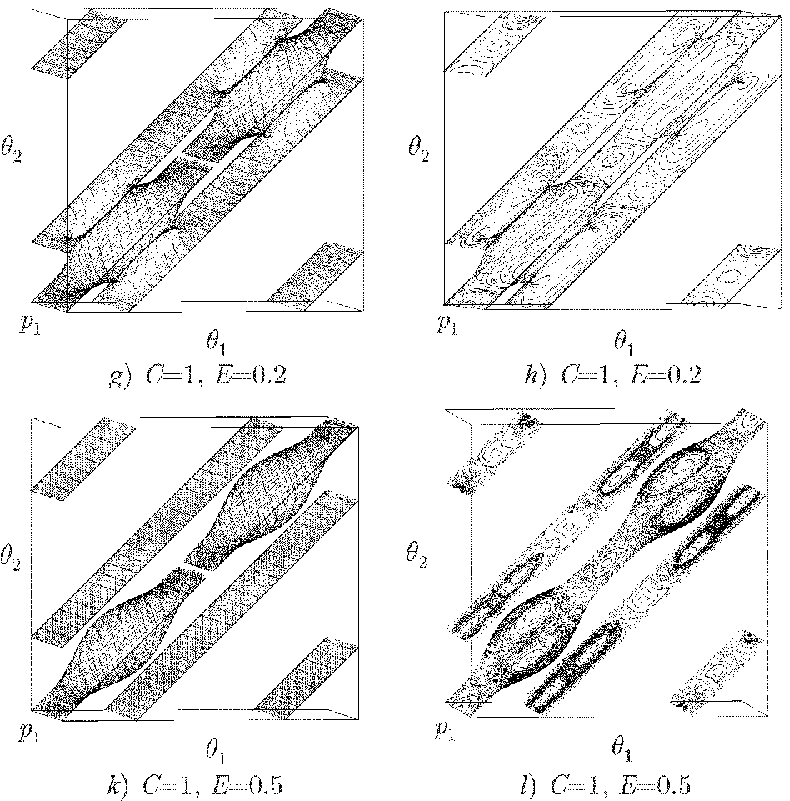}[(Continuation.) Energy level
sections ({\it a, c, e, g, k}\/) and corresponding Poincar\'e maps ({\it
b, d, f, h, l}) for different values of~$E$ and~$c$.]

\subsection{The surface of section. Chaos}

To study the system's behavior numerically, we plot the Poincar\'e section
of the reduced system~\eqref{bmk2-eq-7} in the following way. The
two-dimensional surface of section of the three-dimensional energy
level~$\cH = E$ is given by~$p_2^{} = \const$. Unlike one-and-a-half
degree of freedom systems (or trivial two-degree systems), the surfaces of
section in our case might have a complicated structure. Moreover, the
projection of such a surface on a plane (for
example,~$(p_1^{}$,~$\ta_1^{})$) has singularities. To avoid them, we will
plot these surfaces in the three-dimensional space of
variables~$(\ta_1^{},\,p_1^{},\,\ta_2^{})$. Some examples of surfaces of
section~$p_2^{} = 0$ are given in Fig.~\ref{pic1}. The section's type
depends on the location of the point~$(E,\,c)$ on the bifurcation diagram
(Fig.~\ref{3.eps}). The phase flow of the system~\eqref{bmk2-eq-7}
generates a point Poincar\'e map on these surfaces. The Poincar\'e maps
for the case of equal masses (for some values of~$c$ and~$E$) are shown in
Fig.~\ref{pic1}. According to~\eqref{kor} the points with
coordinates~$(\ta_1^{},\,\ta_2^{},\,p_1^{})$ and~$(\pi - \ta_1^{},\,\pi -
\ta_2^{},\,-p_1^{})$ should, generally, be identified. For large
energies, there are chaotic motions on the sections (Fig.~\ref{pic1}{\it
d}, {\it f}, {\it l}\/), which prevent the existence of an additional
analytic integral of the system~\eqref{bmk2-eq-7}.

\begin{rem*}
Non-existence of an additional analytic integral for~\eqref{bmk2-eq-7} can
also be proved using Poincar\'e's methods~\cite{bmk2-16}, i.\,e. analyzing
the secular terms of the perturbed system and the corresponding number of
non-degenerate periodic orbits. In this connection, the expansions of the
disturbing function, given in~\cite{bmk2-16}, may be useful. These
expansions were used in the proof of non-integrability of the classical
three-body problem.
\end{rem*}

\subsection{Periodic solutions and choreographies}

An important part in the classical celestial mechanics is played by
periodic orbits in a center-of-mass reference frame. Such orbits were
studied in many papers~\cite{lit16}. Recently, a new class of periodic
solutions in the problem of~$n$ equal-mass bodies has been discovered.
These are so-called (\emph{relative\/}) \emph{choreographies\/} --- solutions, when the
particles follow each other (in a rotating frame of reference) along the
same curve, with time shift~$T/n$ (\cite{lit15}, see also~\cite{lit14}).

A similar part in the two-body problem on a sphere is played by periodic
solutions of the reduced system~\eqref{bmk2-eq-7}. The simplest periodic
solution that can be expressed analytically is~\eqref{eq21_5}. There are
other similar solutions, which, however, cannot be presented by analytical
formulas. A suitable tool to analyze these solutions is the Poincar\'e map
constructed above. Fixed points of the least order correspond to the simplest
periodic solutions (see Fig.~\ref{pic1}). These solutions define
choreographies of particles (the particles move along the same curve) in
some rotating frame of reference~\cite{lit14,lit15}. For the least-period
periodic solutions, the projections of such choreographies on a plane,
perpendicular to the axis of rotation, are given in Fig.~\ref{pic2}.
Figures~$A_1$\2$A_6$ and~$B_1$\2$B_3$ show two different stable solutions,
while figures~$C_1$\2$C_2$ show the unstable solution of the reduced
system for different energies. Figures~$B_1^{},\,B_2^{}$, and~$B_3^{}$
show the analytic solution~\eqref{eq21_5}.

\fig<bb=0 0 120.0mm 175.2mm>{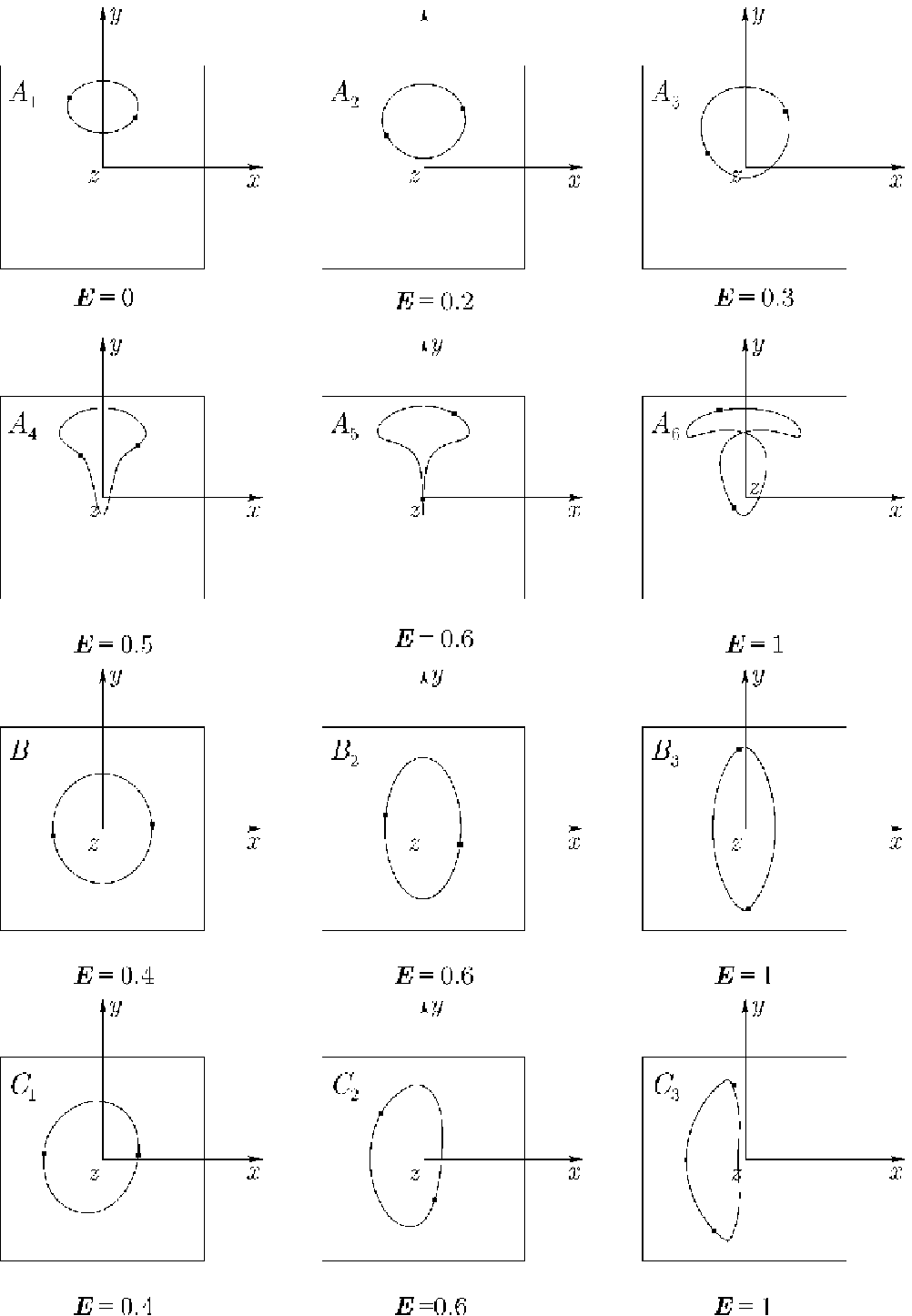}[Relative choreographies for the
problem of two equal-mass bodies ($m_1 = m_2 = 1$) on a sphere,~$c=1$,
$R=1$. \label{pic2}]

\begin{rem*}
It can be shown, using the methods of~\cite{lit14}, that for a
fixed~$c$, there exists a countable set of values of~$E$, for which the
particles form an absolute choreography, i.\,e. move along a closed curve
in a fixed space.
\end{rem*}

\begin{rem*}
When $m_1^{} \ne m_2^{}$, choreographies are broken~--- each particle
moves along its own closed curve (in a rotating frame of reference).
\end{rem*}

\textbf{Acknowledgement}

We gratefully acknowledge the support from the program ``State Support for Leading
Scientific Schools'' (grant~{\selectlanguage{russian}НШ}\136.2003.1), the Russian Foundation for Basic Research
(grant~04\105\164367) and the U.S. Civilian Research and Development Foundation
 (grant~RU-M1-2583-MO-04).

\end{document}